\documentclass[%
 aip,
 amsmath,amssymb,
 reprint,%
]{revtex4-1}

\usepackage{graphicx}
\usepackage{dcolumn}
\usepackage{bm}
\usepackage{subfigure}
\usepackage[utf8]{inputenc}
\usepackage[T1]{fontenc}

\usepackage{mathptmx}
\begin{document}

\title{Designing Few-layer Graphene Schottky Contact Solar Cell: Theoretical Efficiency Limits and Parametric Optimization}
\author{Xin Zhang}
\email{zhangxin0605@jiangnan.edu.cn}
\author{Jicheng Wang}
\affiliation{School of Science, Jiangnan University, Wuxi 214122, China}

\author{Yee Sin Ang} 
\email{yeesin_ang@sutd.edu.sg}

\affiliation{Science, Math and Technology, Singapore University of Technology and Design, Singapore 487372}

\author{Juncheng Guo} 
\affiliation{Deparment of physics, Fuzhou University, Fuzhou 350116, China}

\date{\today}%

\begin{abstract}
We theoretically study the efficiency limits and performance characteristics of few-layer graphene-semiconductor solar cells (FGSCs) based on a Schottky contact device structure. We model and compare the energy conversion efficiency of various configurations by explicitly considering the non-Richardson thermionic emission across few-layer graphene/semiconductor Schottky heterostructures. The calculations reveal that ABA-stacked trilayer graphene-silicon solar cell exhibits a maximal conversion efficiency exceeding 28\% due to a lower reversed saturation current when compared to that of the ABC-stacking configuration. 
The thermal coefficients of PCE for ABA and ABC stacking FGSCs are -0.064\%/K and -0.049\%/K, respectively. 
Our work offers insights for optimal designs of graphene-based solar cells, thus paving a route towards the design of high-performance FGSC for future nanoscale energy converters.
\end{abstract}

\maketitle

Two-dimensional-material-based heterostructures have been actively explored \cite{akinwande2014two,fiori2014electronics,wang2015van,rodriguez2016enhanced} in recent years due to their widespread applications in nanoelectronics, nanophotonics, and optoelectronics, such as energy harvesting \cite{li2010graphene,javadi2020theoretical,zhang2018graphene}, transistors \cite{wang2019schottky,choi2019low,jiang2020black}, photo detection \cite{lopez2013ultrasensitive}, sensors \cite{kim2013chemically}, and data storage devices\cite{li2017floating}. Due to their unusual optical and electronic properties, 2D materials have immense potential for next‐generation solar cells \cite{das2019role,cheng20182d}. With the scaling trends in photovoltaics moving toward thin, atomically-thin 2D materials with high mechanical strength and flexibility have become the key candidate materials for the development of next-generation photovoltaic technology. In terms of solar energy harvesting, 2D-metal/3D-semiconductor solar cells, especially graphene-silicon contact, have been extensively studied \cite{li2015carbon,won2010photovoltaics,ihm2010number,li2014schottky}, with a particularly strong emphasis on improving the power generation efficiency of the device \cite{miao2012high,shi2013colloidal,meng2016interface}. The record PCE of such devices has been rising steadily in recent years, achieving an exceptional values of 15.6\% and 18.5\%, respectively, for graphene/silicon \cite{song2015role} and graphene/GaAs \cite{li201518} solar cells.
\par Despite extensive research focusing on monolayer graphene solar cells, few-layer-graphene/semiconductor solar cells (FGSCs) with a new transport mechanism remain relatively less explored. Correspondingly, the fundamental efficiency limit and performance characterization of FGSCs remain poorly understood. Importantly, FLG is known to possess completely different electrical and optical properties when compared to monolayer graphene \cite{nakamura2008electric,zhu2014optical,mak2010evolution}. The incorporation of FLG with 3D semiconductors has thus led to myriads of interfacial transport and charge injection phenomena that are distinctive from the monolayer counterpart \cite{ang2016current,sinha2014ideal,ang2018universal,trushin2018theory,javadi2019sequentially}. In relevance to solar cell applications, employing FLG as a top-layer material offers the following advantages. Firstly, FLG possesses high transparency in the visible range of the solar spectrum, thus ensuring sunlight transmission into the semiconductor medium with small enough optical loss. Secondly, FLG exhibits a lower reverse saturation current in contrast to conventional metal and monolayer graphene. As the layer number increases, the series resistance and ideality factor in FLG-based Schottky devices are vanishingly small. Thirdly, since the depletion region forms on the right side of the semiconductor surface, FGSCs are characterized by high photocurrent generation and considerable short-wavelength photoresponse. Benefited by these factors, FGSCs hold strong potential as a candidate materials for low-cost and high-efficiency solar-to-electricity energy conversion.

In this Letter, we investigate the various performance metrics and efficiency limits of FGSCs (see Fig. 1) by harnessing the thermionic emission process in FLG-semiconductor Schottky junctions. Here we consider two forms of FLG: ABA and ABC stacking order configurations, which exhibits layer-and stacking-dependent electronic band structures. To gain physical insights into the main limiting factors and further improve the device performance, we perform a computational study to understand how the device performance are influenced by the effect of stacking order, semiconductor bandgap, layer number of FLG, and temperature. Our results offer practical insights on the optimum design of high-performance FGSCs, thus offering an important theoretical basis for the exploration of 2D-material-based solar energy harvesting technology.

\begin{figure}
\centering
\includegraphics[height=6cm]{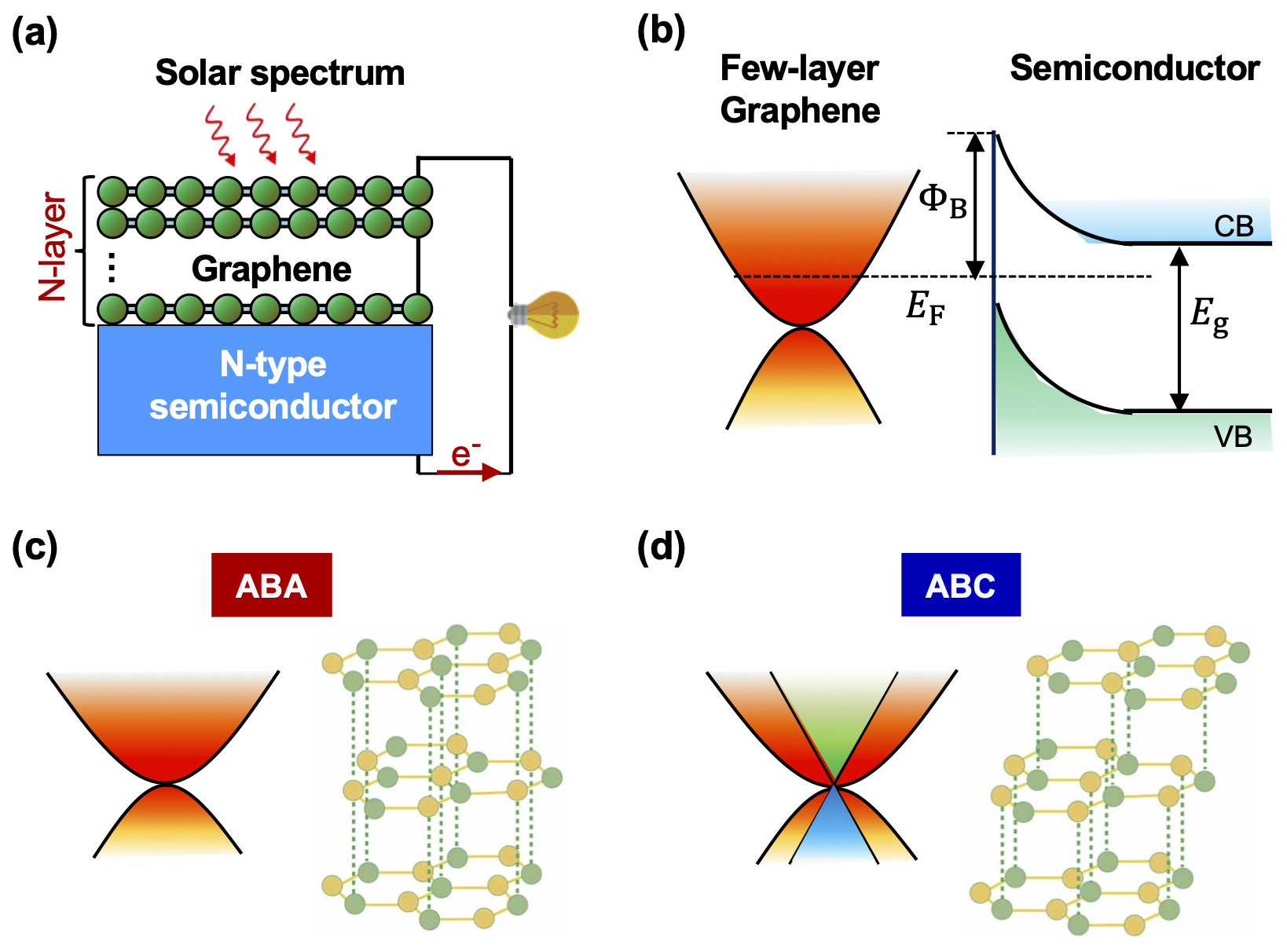}
\caption{(a) Schematic of the vertical few-layer graphene-semiconductor solar cell for sunlight-to-electricity energy conversion. (b) Band diagram showing the thermionic transport over a Schottky barrier $\Phi_{\text{B}}$. (c) ABA- and (d) ABC-stacking-order trilayer graphene with low-energy dispersion (left) and crystal structure (right).}
\label{fig:graph}
\end{figure}

We model an ideal FGSC and determine its theoretical efficiency limit based on the following assumptions: (i) Each incident photon with energy higher than the semiconductor bandgap [$E_{\text{g}}$ in Fig. 1(b)] can produce an electron–hole pair that contribute to the photocurrent, i.e., negligible recombination of excess carriers. (ii) The device ideality factor is $n=1$, and ohmic loss due to the series resistance is expected to be small. Such ideal conditions may be approached practically via FLG because both the series resistance and ideality factor are shown to decrease with the number of graphene layers \cite{jie2013graphene}.
The $J-V$ characteristic of the 2D-material/semiconductor Schottky contact at room temperature exhibits a rectifying behavior as governed by the Shockley diode equation,
\begin{equation}\label{01}
J(V,T)=J_{\text{SC}}-J_{\text{SAT}}\left[\exp\left(\frac{qV}{k_{\text{B}}T}\right)-1\right],
\end{equation}
where $J_{\text{SC}}$ represents the short circuit photocurrent density, $J_{\text{SAT}}$ denotes the reversed saturation current density (RSCD) due to thermionic emission over the Schottky barrier [$\Phi_{\text{B}}$, see Fig. 1(b)] under a bias voltage $V$, $q$ is the electron charge, $k_{\text{B}}$ is the Boltzmann constant, and $T$ is the operating temperature of the cell. The short-circuit current or photocurrent density generated by sunlight absorption in FSSC is
\begin{equation}\label{02}
J_{\text{SC}}=q\int_{E_{\text{g}}}^{\infty} \mathcal{T}(\hbar\omega)I_{\text{AM1.5D}}(\hbar\omega)\, d\omega,
\end{equation}
where $I_{\text{AM1.5D}}$ is the incident photon flux of AM1.5D terrestril solar spectrum with wavelength range of 280 nm to 4000 nm. Here $\mathcal{T}$ represents the optical transmittance of FLG, which can be obtained from experimental data \cite{zhu2014optical,mak2010evolution}. $E_{\text{g}}$ is the bandgap of the semiconductor, and $\hbar\omega$ represents the photon energy with frequency $\omega$.
The open-circuit voltage can then be obtained as
\begin{equation}\label{01}
V_{\text{OC}}=k_{\text{B}}T\ln(J_{\text{SC}}/J_{\text{SAT}}+1)/q.
\end{equation}
For an ideal Schottky contact, the transport of charge carriers is governed by the thermionic emission over the Schottky barrier at the contact interface and the corresponding RSCD is given by a generalized thermionic emission model
\begin{equation}\label{01}
J_{\text{SAT}}=\mathcal{A}T^{\beta}\exp(-\Phi_{\text{B}}/k_{\text{B}}T),
\end{equation}
where the prefactor $\mathcal{A}$ and the scaling exponent $\beta$ are material-and interface-dependent parameters \cite{ang2018universal}. The scaling exponent takes the form of $\beta=2$ and $\mathcal{A}=120$ A/cm$^{2}$K for a classic Schottky contact composed of 3D bulk metals with parabolic energy dispersion. 
For FLG-based vertical Schottky contact, the electronic properties of the FLG \cite{ang2016current,sinha2014ideal,ang2018universal,trushin2018theory,javadi2019sequentially}, deviates significantly from the parabolic energy dispersion. 
Furthermore, the electronic properties of FLG exhibit nontrivial dependences on the number of layers and the layer stacking order. 
For instance, the bernal (i.e. the ABA stacking order) and the rhombohedral (i.e. the ABC stacking order) FLG -- the most thermodynamically stable stacking orders \cite{neto2009electronic} [see Fig. 1(c) and (d)] -- display highly nonparabolic energy dispersions. 
Such band structure \emph{nonparabolicity} gives rise to an unconventional and non-Richardson RSCD whose temperature scaling behavior is drastically different from that of the classic Richardson thermionic emission model. 
In this case, the vertical FLG-semiconductor Schottky heterostructures obeys a current-temperature scaling exponent of $\beta=1$, with the following layer number $N$ and stacking order dependent prefactor \cite{ang2018universal}:
\begin{equation}
\mathcal{A}=
\left\{
\begin{array}{lr}
            \frac{2}{\pi}\frac{Nq}{\tau_{\text{inj}}}\frac{k_{\text{B}}\Phi_{\text{B}}}{(\hbar v_{\text{F}})^2} &\text{(ABA)}\vspace{1ex}  \\
             \frac{2}{\pi}\frac{q}{N\tau_{\text{inj}}} \frac{k_{\text{B}}\Phi_{\text{B}}^{2/N-1}}{(\hbar v_{\text{F}})^2}  &\text{(ABC)} \\
             \end{array}
,\right.
\end{equation}
where $N$ represents the FLG layer number, $\tau_{\text{inj}}$ is a charge injection characteristic time constant whose value is influenced by the quality of the contact, $\hbar$ is the reduced Plack's constant, and $v_{\text{F}}=10^6$ m/s. Here, the RSCD shows an $N$-fold enhancement in ABA FLG due to the presence of $N$ conduction subbands. This is in stark contrast to ABC FLG where the RSCD has a nonlinear $N$-dependent prefactor. 

The PCE of the FGSC is defined as the ratio of the maximum electric power to the total incoming solar photon energy flux $P_{\text{sun}}$, i.e.
\begin{equation}\label{02}
\eta=\frac{(JV)_{\text{max}}}{P_{\text{sun}}}=\frac{J_{\text{SC}} \cdot V_{\text{OC}}\cdot {FF}}{\int I_{\text{AM1.5D}}(\hbar\omega)\, d\omega},
\end{equation}
where $FF$ is the fill factor. The maximum power point $(JV)_{\text{max}}$ can be determined by solving $d(JV)/dV=0$.

For a given semiconductor, the PCE increases with the Schottky barrier height as indicated in Eq. (4). Taking the semiconductor bandgap as the ultimate boundary of the Schottky barrier height, i.e., $E_{\text{g}}=\Phi_{\text{B,max}}$, we obtain the upper limits of the PCE for the FGSC. We choose a relatively larger injection time of $\tau_{\text{inj}}=40$ ps to represent the inevitable presence of defects at the Schottky contact interface. 
Figure 2(a) shows the PCE of the FGSC as a function of the semiconductor bandgap at room temperature. The small oscillations originate from the atmospheric absorption in the incident AM1.5D solar spectrum. The efficiency bounds of the FGSC reveal a broad range from 0.8 to 1.5 eV with PCE exceeding 25\%. 
For narrow-bandgap semiconductors, although a high photocurrent density is warranted, the output voltage remains low due to the limited Schottky barrier height (or bandgap) as expected from Eq. (3). On the other hand, for wide-bandgap semiconductors, the photon absorption is significantly impeded by the larger bandgap, which leads to a low photocurrent density [see Fig. 2(b) and Eq. (2)]. The interplay between these two counteracting behaviors leads to an optimal semiconductor bandgap for achieving maximum PEC. The open-circuit voltage is associated with $\mathcal{A}$, and is thus sensitively influenced by the staking order. Importantly, the open-circuit voltage of the ABA-trilayer FGSC is higher than that of the ABC FGSC. The stacking order dependence becomes particularly more pronounced at the wide-bandgap regime and causes an enhancement of the fill factor.

\begin{figure}[htbp]
\centering
\subfigure{
\includegraphics[width=3.95cm]{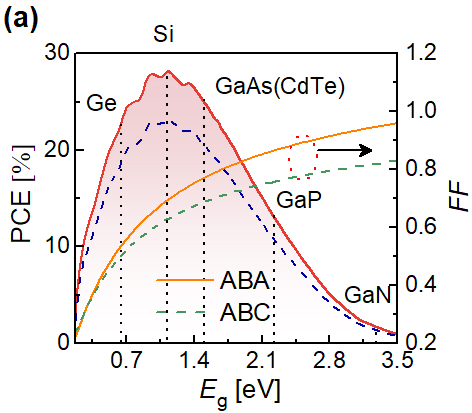}
}
\quad
\subfigure{
\includegraphics[width=3.95cm]{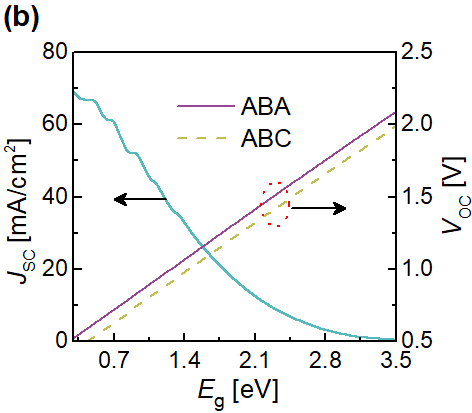}
}
\caption{The performance characteristics of the ABA and ABC trilayer graphene-semiconductor solar cell with carrier injection time $\tau_{\text{inj}}=40$ ps at $300$ K. (a) The PCE ($\eta$) and fill factor ($FF$), (b) short-circuit current density ($J_{\text{SC}}$) and open-circuit voltage ($V_{\text{OC}}$) varying with the semiconductor bandgap ($E_{\text{g}}$). Solid (dashed) lines denote ABA (ABC) stacking.}
\label{fig:graph}
\end{figure}

As shown in Fig. 2(a), the theoretical upper limit of the PCE for the FGSC with ABA trilayer graphene is predicted to be 28.2\% at $E_{\text{g}}=1.12$ eV -- a value which is very close to the bandgap of silicon. To explain why ABA stacking order reflects better performance than ABC, we choose the trilayer graphene/Si solar cell as an illustrative example and further access its $J-V$ characteristics under different operating conditions (see Fig. 3). 
We show that the open-circuit voltage of ABA FGSCs (0.927 V) is higher than that of ABC FGSCs (0.838 eV), although both of them exhibit the same short-circuit current density of 43.8 mA/cm$^2$. These facts demonstrate the key advantage of the ABA-stacking order in achieving high-performance Schottky-contact based solar energy converter. It is worth mentioning that the classic Richardson thermionic emission model, which does not accurately capture the reduced dimensionality and the nonparabolic nature of the energy dispersion in FLG, can yield a discrepancy in the thermionic current density of about 2 orders of magnitude \cite{ang2016current}. In relevance to the modeling of FGSC, using the classic Richardson model of $\mathcal{A}=120$ A/cm$^2$ K$^2$ and $\beta=2$ can lead to an overly small open-circuit voltage as well as a low PCE ($V_{\text{OC}}=0.615$ V and PCE=18.6\%). 
This finding highlights the importance of utilizing the appropriate thermionic emission that better captures the material properties of FLG during the modeling of Schottky-contact based energy converters.

\begin{figure}
\centering
\includegraphics[height=3.8cm]{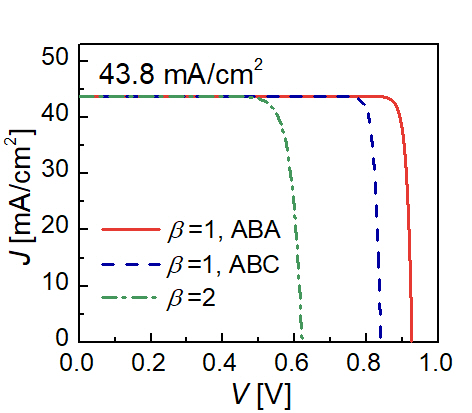}
\caption{$J-V$ characteristics of trilayer graphene-silicon solar cells with ABA and ABC stacking for $\beta=1$ scaling law and classical Richardson–Dushman scaling law ($\beta=2, \mathcal{A}=120$ A/cm$^2$ K$^2$) at 300 K.}
\label{fig:graph}
\end{figure}

Figure 4(a-d) shows the PEC, fill factor, short-circuit current density, and open-circuit voltage of the ABA and ABC FGSC as a function of the layer number of graphene. 
We consider $N\leq6$ because the work function of FLG with more than 6 layers tends to saturate at that of graphite. We identify an optimal layer number of $N=3$ that yields the maximum PCE. 
When the layer number is increased from $N=1$ to $N=3$, the fill factor incraeses substantially. However, as the layer number is further increased, the FF becomes saturated [Fig. 4(b)]. 
In contrast, the short-circuit current density and the open-circuit voltage exhibit a monotonous decreasing trend with the layer number Fig. 4(c) [Fig. 4(d)]. 
The trade-off between the fill factor, short-circuit current density, and the open-circuit voltage leads to an optimal layer number maximizing the PCE. 
To compare the RSCD of FLG with ABA and ABC stacking, we define the ratio of $J_{\text{SAT}}^{\text{(ABC)}}/J_{\text{SAT}}^{\text{(ABA)}}$ and plot the layer dependence of this ratio in Fig. 4(c). 
The $J_{\text{SAT}}^{\text{(ABC)}}$ is about an order of magnitude larger than $J_{\text{SAT}}^{\text{(ABA)}}$, which reveals the fundamental origin of the better ABA FGSC performance in providing a larger open-circuit voltage and a higher PCE when compared to that of the ABC FGSC.

\begin{figure}
\centering
\includegraphics[height=6.5cm]{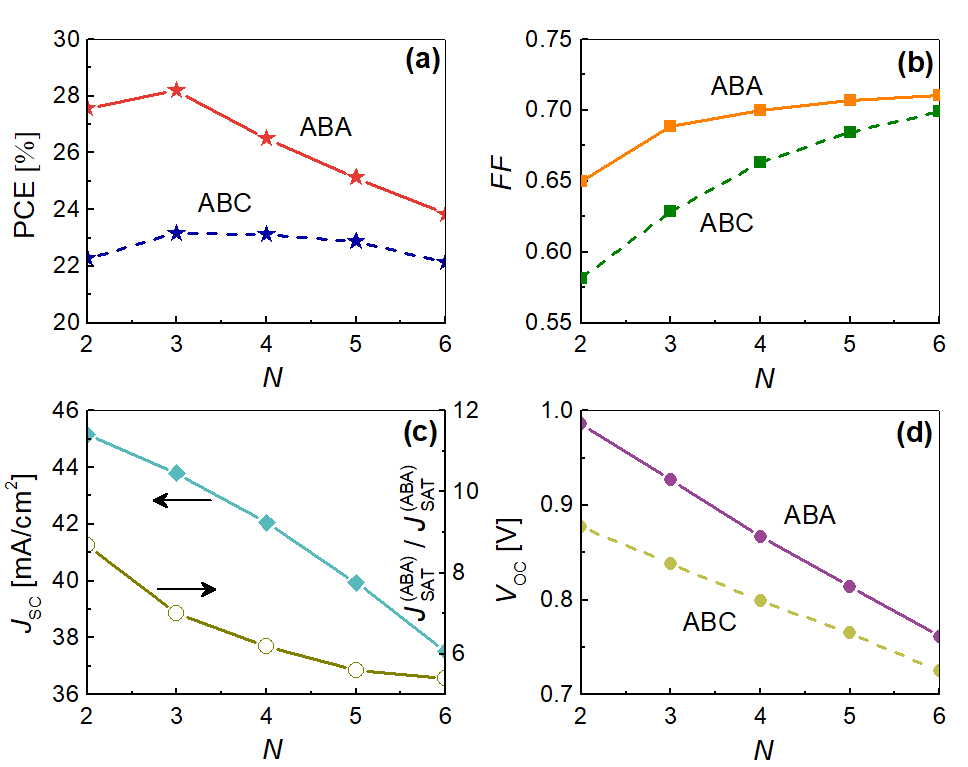}
\caption{The PCE (a), fill factor (b), short-circuit current density (c), and open-circuit voltage (d) of graphene-silicon solar cells with ABA and ABC stacking as a function of graphene layer number at 300 K. Solid (dashed) lines denote ABA (ABC) stacking.}
\label{fig:graph}
\end{figure}

We further study the $\tau_{\text{inj}}$-dependence of the PCE, fill factor, and the open-circuit voltage in Figs. 5(a-c). The $\tau_{\text{inj}}$ is related to the contact quality between the FLG and the semiconductor. Here we consider a representative range from 0.1 ps to 100 ps, which is consistent with the values reported experimentally \cite{sinha2014ideal,xia2011origins,massicotte2016photo,trushin2018theory}. 
A larger $\tau_{\text{inj}}$ corresponds to the situation in which the contact resistance across the Schottky contact is large \cite{sinha2014ideal}. In general, the key performance parameters of the FGSC is influenced by the values of $\tau_{\text{inj}}$. 
Particularly when $\tau_{\text{inj}}$ is small, increasing $\tau_{\text{inj}}$ leads to a significant improvement of the PCE, fill factor, and the open-circuit voltage [see Fig. 5(a-c)]. Such improvement eventually saturates as $\tau_{\text{inj}}$ is further increased. This analysis thus suggests that electrical contact engineering may offer a route to improve the system performance of FGSCs.

Since the RSCD is exponentially dependent on the temperature, the performance of FGSC exhibits a strong temperature dependence as shown in Fig. 5(d-f). 
A high operating temperatures decrease the fill factor [Fig. 5(e)] and open-circuit voltage [Fig. 5(f)], which eventually degrade the PCE [Fig. 5(d)]. Reducing the operating temperature is thus crucially important to achieve optimal energy conversion efficiency. 
For the ABA FGSC, the upper limit of the PCE increases (decreases) from 28.2\% at room temperature to 34.4\% (21.6\%) at $T=200$ (400) K. Similarly, the maximum PCE of the ABC solar cell increases (decreases) from 23.1\% at room temperature to 28.4\% (18.7\%) at $T=200$ (400) K. For both stacking orders, the efficiency limit of the FGSC acts as a nearly perfect linear function of the operating temperature, thus allowing us to define a \emph{thermal coefficient} of the PCE as $\gamma_{\text{PCE}}=\Delta\eta/\Delta T$. For ABA trilayer graphene/silicon solar cells, we obtain $\gamma_{\text{PCE}}^{\text{ABA}}=-0.064$ \%/K, signifying a drop of $0.064$\% in the PCE when the temperature is increased by 1 K. In the case of ABC, the thermal coefficient of PCE is found to be $\gamma_{\text{PCE}}^{\text{ABC}}=-0.049$ \%/K. In addition, we also obtain the thermal coefficient of the fill factor and open-circuit voltage, as depicted in Fig. 5(e) and Fig. 5(f). 

\begin{figure}[htbp]
\centering
\subfigure{
\includegraphics[width=3.85cm]{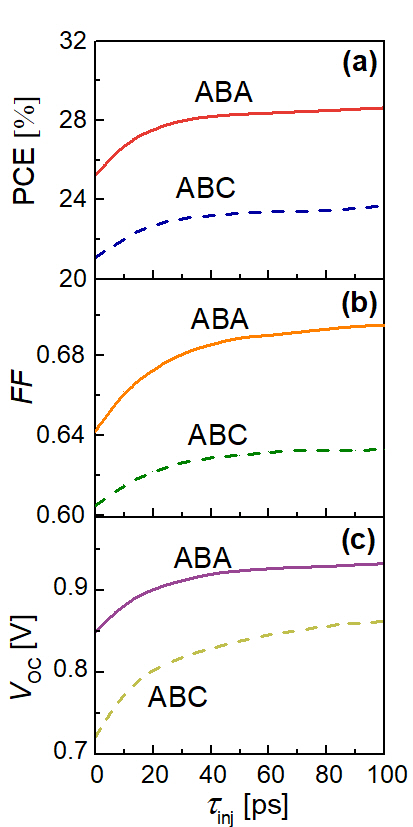}
}
\quad
\subfigure{
\includegraphics[width=3.85cm]{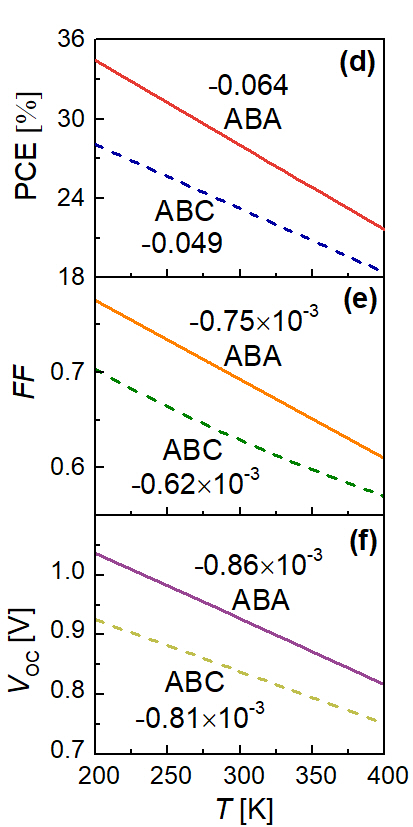}
}
\caption{The carrier injection time dependence (a-c) and temperature dependence (d-f) of the PCE, fill factor, and open-circuit voltage in the FGSC. Solid (dashed) lines denote ABA (ABC) stacking.}
\label{fig:graph}
\end{figure}

In summary, we have performed a computational modeling on the design of FGSC for solar energy harvesting. 
The FGSC with ABA trilayer graphene/silicon architectures possess a peak efficiency of 28.2\% at room temperature, which is significantly higher than that of ABC-FLG-based devices. The better performance of ABA FLG in FGSC application originates from the lower RSCD. Importantly, an optimal layer number of $N=3$ is predicted. Our analysis further reveal an intriguing figure of merits, i.e. the thermal coefficient of the PCE, which allows the temperature dependence of the energy conversion performance to be accessed. Our findings shall provide practical insights useful for the design of high-performance FGSCs, thus paving a potential new avenue towards 2D-material-based solar energy harvesting technology approaching the Shockley–Queisser limit.

This project is supported by the National Science Foundation through grant number 11811530052. YSA acknowledge the supports of Singapore MOE Tier 2 Grant (2018-T2-1-007).

\section*{data availability}
The data that support the findings of this study are available from the corresponding author upon reasonable request.

\providecommand{\noopsort}[1]{}\providecommand{\singleletter}[1]{#1}%

\end{document}